\documentclass{article}
\usepackage{ijcai26}

\usepackage{times}
\usepackage{soul}
\usepackage{url}
\usepackage[hidelinks]{hyperref}
\usepackage[utf8]{inputenc}
\usepackage[small]{caption}
\usepackage{graphicx}
\usepackage{amsmath}
\usepackage{amsthm}
\usepackage{booktabs}
\usepackage{algorithm}
\usepackage{algorithmic}
\usepackage[switch]{lineno}

\usepackage{xcolor}
\usepackage{subcaption}
\usepackage{makecell}
\title{Agents Trusting Agents? Restoring Lost Capabilities with Inclusive Healthcare}
\author{
Alba Aguilera$^{1,2}$
\and
Georgina Curto$^{2}$
\and
Nardine Osman$^{1}$
\and
Ahmed Al-Awah$^{3}$
\affiliations
$^{1}$Artificial Intelligence Research Institute (IIIA-CSIC)\\
$^{2}$United Nations University Institute in Macau (UNU)\\
$^{3}$United Nations Economic and Social Commission for Western Asia (UN-ESCWA)\\
\emails
aaguilera@iiia.csic.es,
curto@unu.edu,
nardine@iiia.csic.es,
al-awah@un.org
}

\begin{document}

\maketitle

\begin{abstract}
Agent-based simulations have an untapped potential to inform social policies on urgent human development challenges in a non-invasive way, before these are implemented in real-world populations. This paper responds to the request from non-profit and governmental organizations to evaluate policies under discussion to improve equity in health care services for people experiencing homelessness (PEH) in the city of Barcelona. With this goal, we integrate the conceptual framework of the capability approach (CA), which is explicitly designed to promote and assess human well-being, to model and evaluate the behavior of agents who represent PEH and social workers. We define a reinforcement learning environment where agents aim to restore their central human capabilities, under existing environmental and legal constraints. We use Bayesian calibration of profile-dependent behavioral parameters of PEH agents, modeling the degree of trust and resulting engagement with social workers, which is reportedly a key element for the success of the policies in scope. Our results open a path to mitigate health inequity by building relationships of trust between social service workers and PEH. 
\end{abstract}

\section{Introduction}

A global report published by the World Health Organization (WHO)~\cite{WHO_2023_DigitalHealth} highlights that the underlying causes of ill health often stem from factors beyond the health sector, such as lack of quality housing, education and job opportunities (i.e. the social determinants of health). Homelessness is directly associated with lower life expectancy and higher morbidity compared to the general population. People experiencing homelessness (PEH) suffer from multidimensional health conditions, often including chronic pathologies, infectious diseases, mental health disorders and substance abuse~\cite{lahiguera2022analisis}. 

Homelessness is a worldwide challenge, which has increased in the majority of developed countries \cite{oecd2024homelessness}. The city of Barcelona is no exception. According to \textit{Salut Sense Llar}, an organization of medical doctors specialized in treating PEH, and \textit{Arrels Fundació}, a non-profit working on the alleviation of homelessness, some of the health inequity challenges that PEH face include a systemic exclusion from primary healthcare (PHC), pharmaceutical poverty and a lack of post-discharge assistance~\cite{diagnosi2022barcelona,saumell2024atencion}.  
Having access to PHC in the city of Barcelona requires obtaining the status of registered resident (``\textit{empadronado}") with a permanent address. Even though PEH usually do not have a permanent address, they can obtain the registered resident status by undertaking a process called social registration (``\textit{empadronamiento social}") with the city's social services. 
However, based on the domain-expert inputs, a very significant number of PEH remain non-registered, do not have access to PHC and express a lack of trust towards social workers, often linked to unrelated previous life traumatic experiences. As a result, their health deteriorates over time, often reaching a critical point where they require emergency care and hospitalization (universally provided in the city of Barcelona). Several studies show that if PEH receive PHC services in time, the diagnosis and treatment of their chronic diseases improves significantly, while reducing visits to emergency services and hospital admissions~\cite{joyce2009,otoole2010,ponka2020}. The inequity in receiving PHC has an important impact on the personal suffering and dignity of PEH. In addition, improving equity in PHC has a broader social impact, since emergency services are much more expensive than PHC regular visits. 

\begin{table*}[ht!]
\small
\centering
\renewcommand{\arraystretch}{1.2}
\begin{tabular}{p{2.6cm} p{13.9cm}}
\hline
\textbf{Life} & 
\textbf{Being able to live to the end of one’s lifespan without premature death} \\
\textbf{Bodily health} & 
\textbf{Being in good physical health, being adequately sheltered and being adequately nourished} \\
\textbf{Bodily integrity} & 
\textbf{Being able to move freely, being free from violence, having bodily, reproductive, and sexual autonomy} \\
Senses, imagination, thought & 
Being able to reason, think, and create; having access to art, literature, and science; and enjoying pleasurable experiences while avoiding non-beneficial pain \\
Emotions & 
Being able to form and mourn emotional attachments to others \\
\textbf{Practical reason} & 
\textbf{Being able to conceptualize what is good and plan one’s future} \\
\textbf{Affiliation} & 
\textbf{ (A) Being able to live with and towards others, to engage in various forms of social interaction, etc. (this entails protecting institutions that nourish such forms of affiliation and protecting the freedom of assembly and political speech) (B) Being able to be treated as a dignified being whose worth is equal to others (this entails provisions of nondiscrimination on the basis of race, sex, ethnicity, etc.). } \\ %Subdivided into interactions with others and dignified, non-discriminatory participation in society
Other species & 
Being able to live with concern for animals, plants, and the natural world \\
Play & 
Laughter, play, and recreational activities \\
Control over one’s environment & 
Subdivided into political participation and material rights to own property and undertake employment \\
\bottomrule
\end{tabular}
\caption{Central Capabilities, adapted from~\protect\cite{nussbaum2011}. Targeted capabilities in the proof of concept are marked in bold.}
\label{tab:nussbaum_capabilities}
\end{table*}

This paper responds to the needs expressed by the above mentioned non-profit organizations to evaluate the policies currently under discussion in the city of Barcelona~\cite{Llei212010} on equitable PHC to PEH.  With this goal, we present an agent-based model (ABM) framework and simulate the policy implementation. We engineer and shape a profile-dependent reward function that represents PEH's motivations, as well as the degree of trust towards social workers and institutions, which constitutes a key policy success factor based on the inputs provided domain expert-knowledge and existing literature giving a voice to PEHs ~\cite{WorldBankVoices2000}. The paper presents the first operationalization of the Capability Approach for human development \cite{sen1999,nussbaum2011} in an agent-simulation for policymaking, which allows to model key aspects of human wellbeing such as affiliation (necessary to establish relationships of trust between PEH and social workers). Unlike utilitarian development models~\cite{little1974project}, which focus on maximizing the overall well-being of the population, often through cost-benefit analysis~\cite{layard2005happiness}, the CA shifts the focus to the real opportunities (or capabilities) individuals have to live the life they value \cite{sen1999}. 

\section{Related Work}
We review four key areas of prior research: ABMs for policy-making, ABMs that rely on the CA, RL for modeling human-like decision-making, and calibration of ABMs and RL-driven behavior.  %\textcolor{red}{ABM's calibration, RL-based behavior calibration or both? }. 
\label{sec:relwork}
Agent-based simulations have been widely used as tools to support policy-making in complex social challenges, such as disease outbreaks~\cite{dignum2021social}, gentrification~\cite{eckerd2019gentrification}, spatial inequality~\cite{tomasiello2020access} or urban ageing~\cite{ma2016urbanageing}. They allow policy-makers to explore "what if" scenarios for urgent human challenges, such as health inequity among vulnerable populations, in a non-invasive way (before affecting real-world populations). However, existing ABM applications in contexts of social inequity often oversimplify agents' motivations and overlook structural barriers that leave vulnerable groups of people behind \cite{aguilera2025agent}. This article aims to address limitations by explicitly incorporating the CA as a conceptual framework for social simulations. 

While there is an important body of ABM literature addressing equity and fairness~\cite{williams2022integrating}, few studies explicitly use the CA as a conceptual framework. Existing studies have primarily focused on applications for specific domains, such as energy justice~\cite{melin2021energy,assa2020can,de2020conflicted} and community resilience~\cite{markhvida2020quantification,silva2022commuter,tseng2024ci}. These works mainly examine how measurable magnitudes in the evaluative space (such as resources or wealth) are affected in different scenarios, where behavior is modelled following a utilitarian approach (i.e. maximization of their resources and/or capabilities). We argue that this simplified modeling of behavior does not capture the underlying constraints and motivations influencing people's opportunities and outcomes. 

RL has been widely used to develop agents that behave based on complex intrinsic motivations, such as curiosity~\cite{pathak2017curiosity}, empowerment~\cite{klyubin2005empowerment}, risk awareness~\cite{jarne2025emergent} or even social influences from other agents~\cite{jaques2019social}. RL-based agents learn by interacting with their environment and receiving rewards, similar to how humans learn in behavioral theories. Extensive literature exists on using RL to train autonomous agents to perform tasks or behave desirably in robotics, autonomous driving, game playing, etc.~\cite{sutton2018reinforcement}. However, relatively few studies focus on using RL to approximate reward shaping in social simulations for real-world scenarios. 

Another fundamental challenge in the ABM domain is the validation of the simulations to reproduce observed real-world patterns, since this involves calibrating numerous parameters. This is particularly challenging when parameters drive both individual and system-level outcomes. Likelihood-based approaches to Bayesian estimation have proven to perform well in ABMs~\cite{platt2020comparison} and RL studies~\cite{daw2009trial}. However, to our knowledge, there is limited work providing an integrated calibration that simultaneously targets (i) system-level ABM outputs and (ii) individual-level RL behavior in a single framework.

\begin{figure*}[h!]
    \centering
    \includegraphics[width=0.999\linewidth]{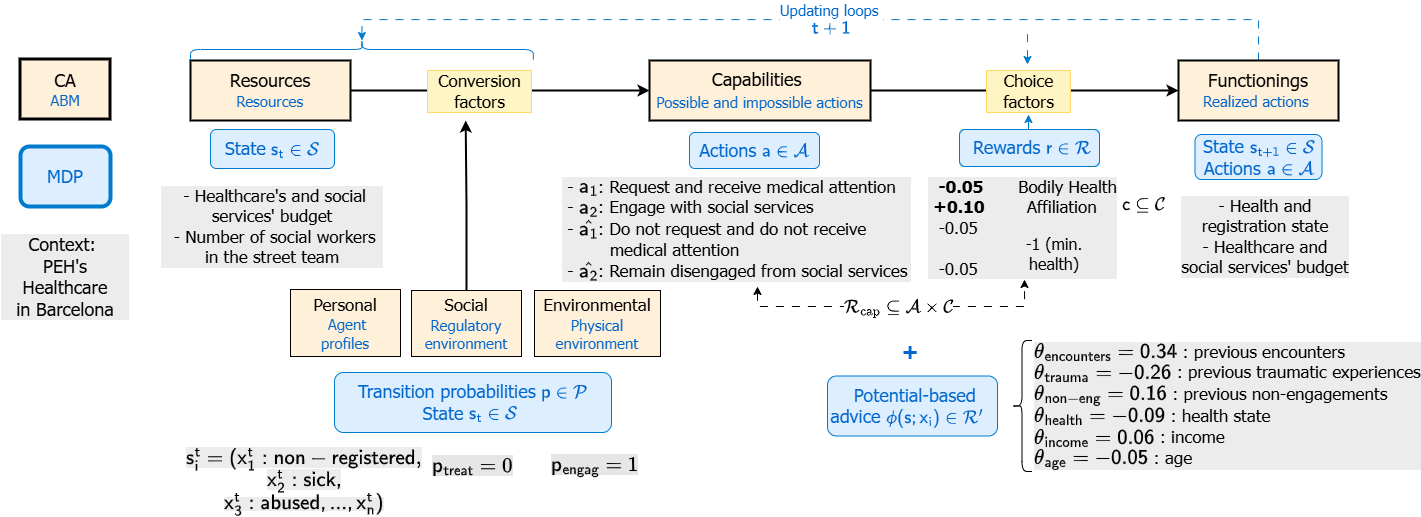}
    \caption{Architecture overview mapping the CA conceptual framework, the ABM and MDP elements and the real-world proof of concept: 1) CA conceptual framework (shown in orange); 2) ABM and MDP elements (shown in blue); 3) the contextualized elements for the proof of concept in Barcelona (shown in grey).}
    \label{fig:scheme}
\end{figure*}

In this context, we claim that existing social simulations, to support non-profit organizations and governments in decision making, fall short in two aspects: (1) modeling complex decision-making processes that reflect diverse stakeholders' motivations and real environmental constraints (while informed by domain expert knowledge), (2) assessing inequity and social justice in terms of individual's real opportunities, among other relevant elements. Our work addresses the first gap by designing profile-dependent RL rewards that capture the agents' motivation to restore their capabilities, calibrated using data and domain expert knowledge inputs, to better approximate stakeholders' behavior in the context of primary health services accessible to PEH. It addresses the second by adopting a CA-aligned, pluralistic evaluation that tracks what agents are effectively able and unable to do. 
\section{Method}

The main objective of this work is to evaluate policies, focusing on the improvement of health services to PEH in Barcelona, through the lens of their impact on individual capabilities and equity. To achieve this, we develop an agent-based simulation that models people's behavior under different policy scenarios.

The simulation contains three main elements: (i) the agent population, (ii) the physical environment and (iii) the regulatory environment. The main type of agent in the simulation represents people experiencing homelessness (PEH), with heterogeneous personal features. PEH agents navigate a physical environment where resources or assistance may be available either through street outreach teams or through social service and healthcare centers. The regulatory environment defines the allocation of these resources across these services, shaping the accessibility, availability and scale of support provided. Our work operationalizes the Capability Approach for human development in an ABM framework ~\cite{aguilera2025agent}, through the agents' decision-making (Section \ref{sec: decision}), the simulation calibration (Section \ref{sec: calibration}) and the evaluation of the policies in scope (Section \ref{subsec:eval}), based on the requirements of the CA. The implementation is publicly available on GitHub~\footnote{\url{https://github.com/albaaguilera/Inclusive-Healthcare}}.

\subsection{The Agents' Decision Making}
\label{sec: decision}
\paragraph{Conceptual Formulation.}The CA provides the main framework upon which we build the decision-making of the agents. It introduces notions including: resources, conversion factors, capabilities, choice factors and functionings. Following Fig.~\ref{fig:scheme}, we align these CA concepts with elements in ABMs and Markov decision process (MDP) environments, as we illustrate below: 
\begin{itemize}
    \item \textbf{Resources}, which include the set of commodities and services available to a person in a given context (such as public healthcare or social street teams), are represented within the MDP states.
    \item \textbf{Conversion factors} include the personal, social and environmental contexts defining how can we transform resources into capabilities (such as the personal health or registration state, policies defining who gets access to public healthcare, and proximity to social service assistance). These are represented within the agent profiles, the regulatory environment and the physical environment, both in the state and transition probabilities of an MDP. %\textcolor{purple}{I think we should introduce something here about trust}
    \item \textbf{Capabilities} include \textit{what people are able to do and be}, given their resources and conversion factors. They are represented as agents' available actions.
    \item \textbf{Choice factors} include the motivators of behavior (such as human values, needs, emotions) influencing the individual's choice of prioritizing one action over another. In this paper, lack of trust acts as a major choice factor. These are represented as reward functions.
    \item \textbf{Functionings} include\textit{ what people end up doing and being}. These are represented as the actions realized by the agent, which are deterministically tied to the next state of the MDP in this paper.
\end{itemize}
\begin{table*}[h!]
{\small
    \centering
    \renewcommand{\arraystretch}{0.92} % default is 1.0

    \begin{tabular}{p{3.0cm}|p{3.7cm}|p{5.9cm}|p{3.3cm}}%|p{1.0cm}|p{1.3cm}}
        \hline
        \textbf{PEH's actions (possible or impossible)} & \textbf{Domain-expert knowledge that justifies set of actions and rewards} & \textbf{Action implementation in the simulation (feasibility and consequences)} & \textbf{Central capabilities with evaluation weights ($\alpha_k$) and rewards ($r (c_j)$)} \\ \hline 
        $a_1:$ Request and receive PHC attention &
Preventive medical care can reduce visits to emergency services and hospital admissions. Providing inclusive PHC for PEH depends on the legal policies implemented and resources available. &
Trust-reliant action. Feasibility depends on the policy implemented. If possible, health state increases $0.5$ and the agent is moved to the PHC location; a cost of 30 is applied to the healthcare budget. If impossible, health state decreases $0.5$. &
\makecell[tl]{\textbf{Bodily Health} ($\alpha_1 = 0.8$), \\ \textbf{Affiliation} ($\alpha_1= 0.2$), \\ Life, Bodily Integrity,\\
($r=+0.1$ if restored, \\ $r=-0.05$ if deprived).} \\
\hline
$\bar{a}_1:$ Do not request and receive PHC attention &
Contrary to $a_1$. &
Always possible; health state decreases $0.5$. &
\makecell[tl]{Practical Reason \\ (\(r=-0.05\)).} \\
\hline

$a_2:$ Engage with social services &
Engagement with social services is required for PEH to become registered residents. Trust-building is key and engagement is often gradual. &
Trust-reliant action. Feasibility depends on physical adjacency between PEH and social worker agents. If possible, after 2 engagements the agent is moved to social services and registration state is updated. If impossible, health state decreases $0.5$. &
\makecell[tl]{\textbf{Affiliation} \\ ($\alpha_2 = 0.8$ if restored) \\($r=+0.10/-0.05$).} \\
\hline

$\bar{a}_2:$ Remain disengaged from social services &
Contrary to $a_2$. &
Always possible; health state decreases $0.5$. &
\makecell[tl]{Practical Reason \\ (\(r=-0.05\)).} \\
\hline

Receive emergency care / be hospitalized (ICU) &
Emergency visits and hospital admissions are available to everyone but extremely costly, and reflect failure of early healthcare interventions. &
Not an action initiated by the PEH agent. This occurs when the agent reaches a terminal state, where the simulation ends for the agent. A cost of $1000$ is applied to the healthcare budget and the agent is moved to the ICU location. &
\makecell[tl]{\textbf{Bodily Health}, \\ Bodily Integrity, \\ Practical Reason, \\ (\(r=-1\)).} \\
\hline
\end{tabular}
}
    \caption{Summary of the selected PEH's actions $a_k$ (with their contrary $\bar{a_k}$) and their implementation in the simulation. The table also details the mapping between actions and central capabilities, along with the corresponding rewards upon execution and evaluation weights upon policy assessment of the central capabilities in bold. }
    \label{tab:actions_capabilities_rewards}
\end{table*}

In this framework, agents aim to restore or expand deprived capabilities, but may face barriers to do so because of their personal, social and environmental circumstances (conversion factors). Additionally, agents may prioritize to restore one capability over another based on their individual motivators of behavior (choice factors). For instance, a PEH agent may or may not be able to \textit{engage with social services} depending on the accessibility or availability of certain resources (such as the number of social service workers on the street). Additionally, even if this action is possible based on the available resources, it may not be achieved because of choice factors such as the degree of trust towards social workers, associated with certain attributes in the agent's profile (including the number of previous encounters with social workers, duration of homelessness, personal history, including traumatic experiences). 

In Table~\ref{tab:actions_capabilities_rewards}, we present a summary of the selected actions agents perform in our case study. The actions have been identified as key on the topic of PEH and PHC access by the local social workers, healthcare experts and non-profit organizations. The actions are also linked to the CA, as they constitute ways to restore deprived capabilities including affiliation (through the engagement between PEH and social workers) and bodily health (by being treated in PHC). For the sake of clarity, we keep both the list of actions and central capabilities short, though we acknowledge that additional ones should be introduced in future work for a holistic representation and evaluation of this social context. 

\paragraph{Mathematical Formulation. }We formulate the framework as an MDP defined by the tuple $(\mathcal{S}, \mathcal{A}, \mathcal{P}, \mathcal{R})$ where $\mathcal{S}$ and $ \mathcal{A}$ are discrete sets of states and actions, and $\mathcal{P}, \mathcal{R}$ are transition probability and reward functions. States $s\in\mathcal{S}$ store all the information contained in the system at each simulation's timestep $t$, while actions $a\in\mathcal{A}$ represent the options in Table~\ref{tab:actions_capabilities_rewards}, which may be possible or impossible, and desired or avoided, depending on transition probabilities $p \in \mathcal{P}$, and rewards $r \in \mathcal{R}$.

In this section, we define $\mathcal{P}$ and $\mathcal{R}$ to encode (i) how policies and other constraints in the physical and regulatory environment influence agents' behavior, and (ii) how agents prioritize actions based on the restoration of their main capabilities and the nuances derived from their own personal characteristics following domain expert knowledge. For the first one, which involves encoding the conversion factors of the CA, we consider  
\begin{equation}
\small
    \mathcal{P}(s, a, s') = 
    \begin{cases} 
        p, &\text{\footnotesize if $s$ fulfils all requirements}  \\
    1-p, &\text{\footnotesize if $s$ does not fulfil all requirements}
    \end{cases}
\label{eq: transitions}
\end{equation}

where $p$ is the probability of ending up in state $s'$ after executing action $a$, and the requirements involve elements of the agent's state, stored in $s$.

 In our context, the constraints considered in the regulatory and physical environment are completely binary ($p=1$ or $p=0$), and can be encoded in $\mathcal{P}$ using Eq.~(\ref{eq: transitions}) with requirements $s$ = being (or not being) a registered resident, and $s$ = being (or not being) nearby a social service worker, respectively. Therefore, the feasibility of action $a_1$ and $a_2$ is subject to constraints imposed by the registration state and the physical encounters with social worker agents (this defines $p_{\mathrm{treat}}$ and $p_{\mathrm{eng}}$
 in Fig.~\ref{fig:scheme}). In this paper, we focus only on the health and registration state, but the housing state could also be considered with requirement $s$ = having (or not having) available shelter spaces. 

At each simulation's timestep $t$, for a given state $s \in \mathcal{S}$, we keep track of the subset $\mathcal{A}_{pos}(s) \subseteq \mathcal{A}$ of possible actions and the subset $\mathcal{A}_{imp}(s)$ of impossible actions based on the limitations or constraints in the agents' profiles, regulatory and physical environment. These will serve as an indicator for restored and deprived central capabilities in the evaluation. Accordingly, for $a \in \mathcal{A}_{imp}(s_t)$, we set $\mathcal{P}(s, a, s') = 0$ for all $s' \in \mathcal{S}$, ensuring these actions have no effect.

\textit{Reward Engineering. } To define the choice factors of the CA, we engineer a reward function $\mathcal{R}$ encoding agents' motivations. In this paper, the main motivation of the agent is restoring or expanding its central capabilities. We consider a partially ordered set of central capabilities $\mathcal{C} = [c_1, ..., c_n]$ ranked by importance $c_1\succ\dots\succ c_n$. Each action $a \in \mathcal{A}$ can contribute to restoring (or depriving) several capabilities. At the same time, each capability can be restored (or deprived) by more than one action. We capture this binary relation with $\mathcal{R}_\mathrm{cap}\subseteq\mathcal{A}\times\mathcal{C}$, where $(a_i, c_j) \in \mathcal{R}_{cap}$ if action $a_i$ restores (or deprives) capability $c_j$.  Based on these predefined connections between actions and central capabilities informed by domain expert knowledge, one can define the baseline reward of executing an action as the sum of the capability rewards it advances (column five in Table~\ref{tab:actions_capabilities_rewards}):
\begin{equation}
    \mathcal{R}(s, a, s') = 
    \begin{cases} 
    \sum_{(a,c_j)\in\mathcal{R}_\mathrm{cap}} r(c_j) & \text{if } \text{$ s'\notin S_{\mathrm{dep}}$} \\  
    -\rho, & \text{if } \text{$ s'\in S_{\mathrm{dep}}$} 
    \end{cases}
\label{eq: rewards}
\end{equation}

where $S_{\mathrm{dep}}\subseteq\mathcal{S}$ are the set of terminal states with \textit{critical capability deprivation}. In our context, the terminal deprivation states are those where the agent's health reaches its minimum, requiring emergency services and hospitalization (with penalty $\rho = 1$). By defining a large negative reward $\rho\gg r(c_j)$ $\forall j$, we aim to guarantee that the optimal strategy of the agent avoids $S_{\mathrm{dep}}$. 

\textit{Reward Shaping. } To reflect heterogeneity and complexity in the agents' behavior, we introduce profile-dependent variations in the baseline reward function (Eq.~\ref{eq: rewards}). In our context, these variations are supported by the relevant literature describing the difficult choices that characterize behavior below the poverty line ~\cite{nussbaum2011,mullainathan2013scarcity}, as well as by domain expert knowledge from local stakeholders. We introduce this heterogeneity, for each agent $k$, using potential-based reward shaping (PBRS)~\cite{ng1999policy} and its extension to state-action advice potentials~\cite{wiewiora2003principled}. We add a state–action potential $\Phi (s, a)$ that encourages (or discourages) specific actions $a \in \mathcal{A}_{advice}$. The shaped reward is defined as 
\begin{equation}
\begin{aligned}
    \mathcal{R}'(s,a,s') 
    &= \mathcal{R}(s,a,s') \\
    &\quad + \beta\Big(\gamma\,\Phi(s',a';x_k)
    - \Phi(s,a;x_k)\Big).
\end{aligned}
\label{eq:shaped_reward}
\end{equation}
\begin{equation}
\text{with} ~~\Phi (s, a; x_k) = \begin{cases} \theta_{x_k}\,\phi(s,x_k) & \text{if } a \in \mathcal{A}_{advice}~ \\0 & \text{otherwise}  \end{cases} 
\label{eq: qinit}
\end{equation}
where \(\beta \in (0,1)\) controls the strength of the shaping signal. In our context, $\mathcal{A}_{advice}$ only includes trust-reliant actions in Table~\ref{tab:actions_capabilities_rewards}. $\phi(s,x_k)$ encodes the feature expectations over the agent's state and personal attributes $x_k$, either from data or domain expert knowledge (e.g. how a particular number of previous encounters, health state, history of abuse, duration of homelessness, age, gender or income-level may (dis)incentivize certain trust-reliant actions). The parameter vector $\theta$ assigns weights to these features, as illustrated in Fig.~\ref{fig:scheme}. In this way, we can differentiate between two agent clusters: low- and moderate-trust agents. 

\subsection{The Simulation Calibration}
\label{sec: calibration}
Human real-world behavior is heterogenous, and can be simulated in ABM scenarios using available data. Given the limited real-world data on homelessness, in this paper we generate synthetic data by matching macro-level statistics and use it to inform the Bayesian-based calibration methodology. We estimate $\theta$ from data, inject the resulting potential into the simulation, and choose $\beta$, i.e. the strength of the reward shaping signal by comparing simulated system-level statistics to synthetic targets. 

\paragraph{Synthetic behavioral data. } We generate synthetic sequential data representing repeated encounters between PEH agents and social workers on a grid world. We define an encounter as a timestep in which engaging social services ($a_2$) is physically feasible (e.g., the PEH agent is adjacent to a social worker). For each agent $k$, across multiple encounters $e=1, ..., E$, we sample a binary decision between engaging or not engaging $y_{k,e}\in \{0,1\}$ from a sigmoid probability distribution, such that $\mu_{k,e} = \theta^{\top} \phi_{k,e}$ and  
\begin{equation}
p_{k,e}= \sigma(\mu_{k,e}) = \frac{1}{1+\exp(-\mu_{k,e})}, 
\label{eq:synth_sigmoid}
\end{equation}
\begin{equation}
\text{with} ~~ y_{k,e}
\sim \text{Bernoulli}(p_{k,e}),
\label{eq:synth_bernoulli}
\end{equation}

where $\phi_{k,e}$ is a feature vector encoding personal characteristics and $\theta$ is the corresponding vector of coefficients. We use the same feature map defined in Eq.(~\ref{eq: qinit}), i.e. $\phi_{k,e}= \phi(s_{k,e}, x_k)$ at the encounter-level. This produces a demonstration dataset $\mathcal{D}={(\mathbf{x}_{k,e},y_{k,e})}$ from which we compute summary targets: overall engagement rate, engagement rates by trust cluster or registration cluster, and engagement dynamics across repeated encounters.

\paragraph{Inverse calibration of profile-dependent advice.}
We infer $\hat{\theta}$ using Bayesian inverse calibration under the logistic Bernoulli likelihood associated with the binary decision model in Eqs.~(\ref{eq:synth_sigmoid})--(\ref{eq:synth_bernoulli})~\cite{bishop2006pattern,murphy2012machine}.

We choose a Gaussian prior $p(\theta)=\mathcal{N}(\theta;\mu_0,\Sigma_0)$ and compute the posterior
\begin{equation}
p(\theta\mid\mathcal{D})\propto p(\mathcal{D}\mid\theta)\cdot p(\theta).
\label{eq:posterior}
\end{equation}
In practice, we obtain the maximum-a-posteriori (MAP) estimate $\hat{\theta} = \arg\max_{\theta}{p(\theta \mid \mathcal{D})}$ and quantify uncertainty via a Laplace approximation.
This calibration is model-free, since it provides a data-driven parametrization of the potential-based advice without needing to run the simulation itself. We acknowledge that when calibrating against real-world data, the likelihood is typically not so easily available in closed form, and the simulator must be run repeatedly to approximate it, making calibration much more complex and computationally expensive.

\paragraph{Forward simulation with the calibrated behavior. }
We inject the calibrated parameters as potential-based advice following Eq. (\ref{eq: qinit}). We choose $\beta$ by minimizing a loss metric between simulated and synthetic summary statistics. In this way, the individual agent starts learning already biased toward/against some actions, while the simulation successfully matches demonstration data. 

\subsection{The Simulation Assessment}
\label{subsec:eval}
Beyond decision-making, a key contribution of our work lies in how we evaluate the policies through the simulation. By operationalizing the conceptual framework of the CA, we do not aim on redistributing material resources only (which in this case correspond to the access to PHC), but also on the restoration of PEH capabilities, corresponding to the (in)feasibility of actions. 
Following the mapping between actions and central capabilities ($\mathcal{R}_\mathrm{act}\subseteq\mathcal{A}\times\mathcal{C}$), we measure the state of central capabilities like ``bodily health" and ``affiliation" through the actions an agent can or cannot perform. We define the evaluation metric describing a single agent's $i$ central capabilities at each simulation timestep $t$ as 
\begin{equation}
\text{Central Capability}_{i} (t) = \frac{
    \sum_{(a_k, c) \in \mathcal{R}_{\mathrm{act}}} \alpha_k \cdot a_{ik} (t)
}{
    \sum_{(a_k, c) \in \mathcal{R}_{\mathrm{act}}} |\alpha_k|
}
\label{eq: evaluation}
\end{equation}
where the evaluation weight $\alpha_k$ captures the positive (restoring) or negative (depriving) contribution of each action to the central capabilities considered. We set the evaluation weights (listed in Table~\ref{tab:actions_capabilities_rewards}) based on the criteria of domain experts. By using Eq.~(\ref{eq: evaluation}), we are creating a generalizable metric supported by literature~\cite{comim2008measuring} that highlights how measurement should remain context-sensitive, participatory and multidimensional. For the evaluation of central functionings, we focus on the outcomes of the agents, such as their final health and registration state. For a multi-agent scenario, we can numerically average Eq.~(\ref{eq: evaluation}) across all $i = 1, \dots, N$ agents in the simulation. 
\begin{figure*}[t]
  \centering
  \begin{subfigure}[b]{0.5\textwidth}
    \includegraphics[width=\textwidth]{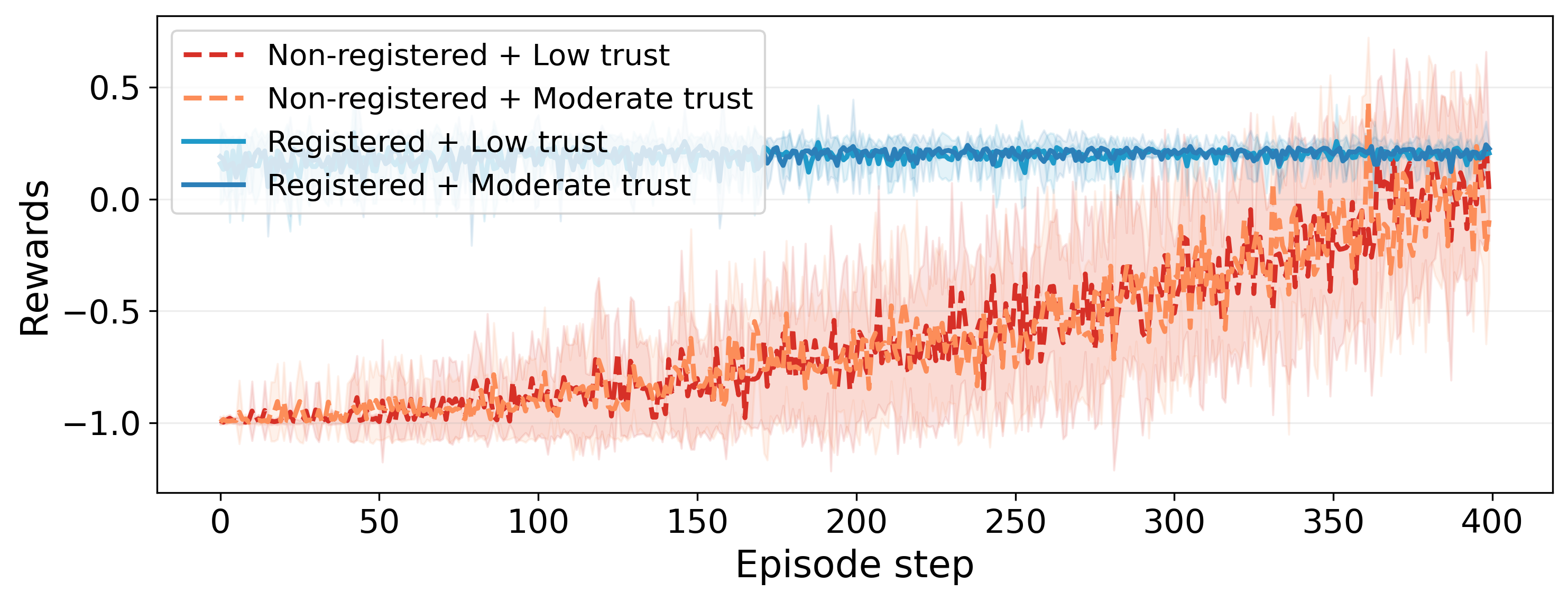}
    \caption{Rewards}
    \label{fig:rewards}
  \end{subfigure}\hfill
  \begin{subfigure}[b]{0.5\textwidth}
    \includegraphics[width=\textwidth]{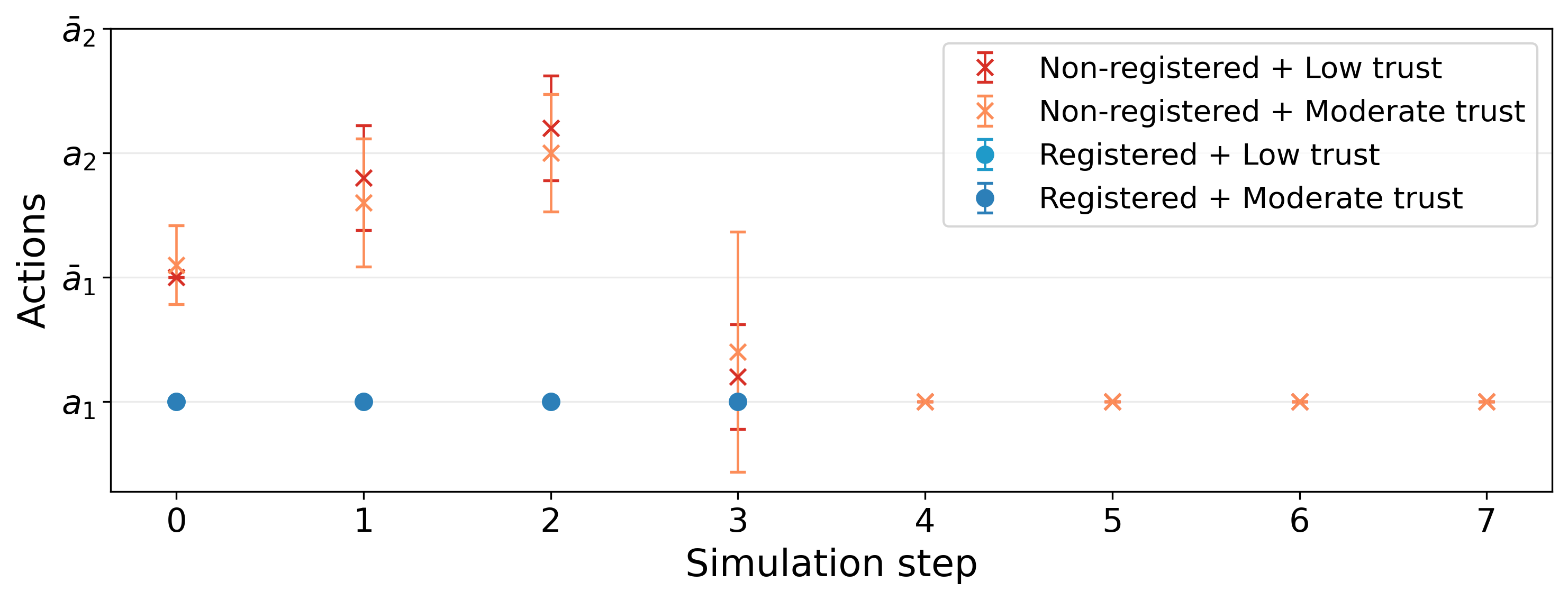}
    \caption{Optimal strategies}
    \label{fig:policies}
  \end{subfigure}
  \caption{Learning dynamics under \emph{policy OFF} and $N=16$ PEH agents across $10$ random seeds: (a) agents' aggregated rewards over episode steps, and (b) mean optimal strategies over simulation steps for the registered (solid lines or dots), non-registered (dashed lines or crosses), low-trust (warm colors) and moderate-trust agents (blue colors), with standard deviation across seeds. } % with trust $\lambda=1$ (in blue and orange) and $\lambda=0.8$ (NOT IMPLEMENTED YET) depending on the agent's profile.}}
  \label{fig:learning}
\end{figure*}
\begin{figure*}[h!]
    \centering
     \begin{subfigure}[b]{0.5\textwidth}
    \includegraphics[width=\textwidth]{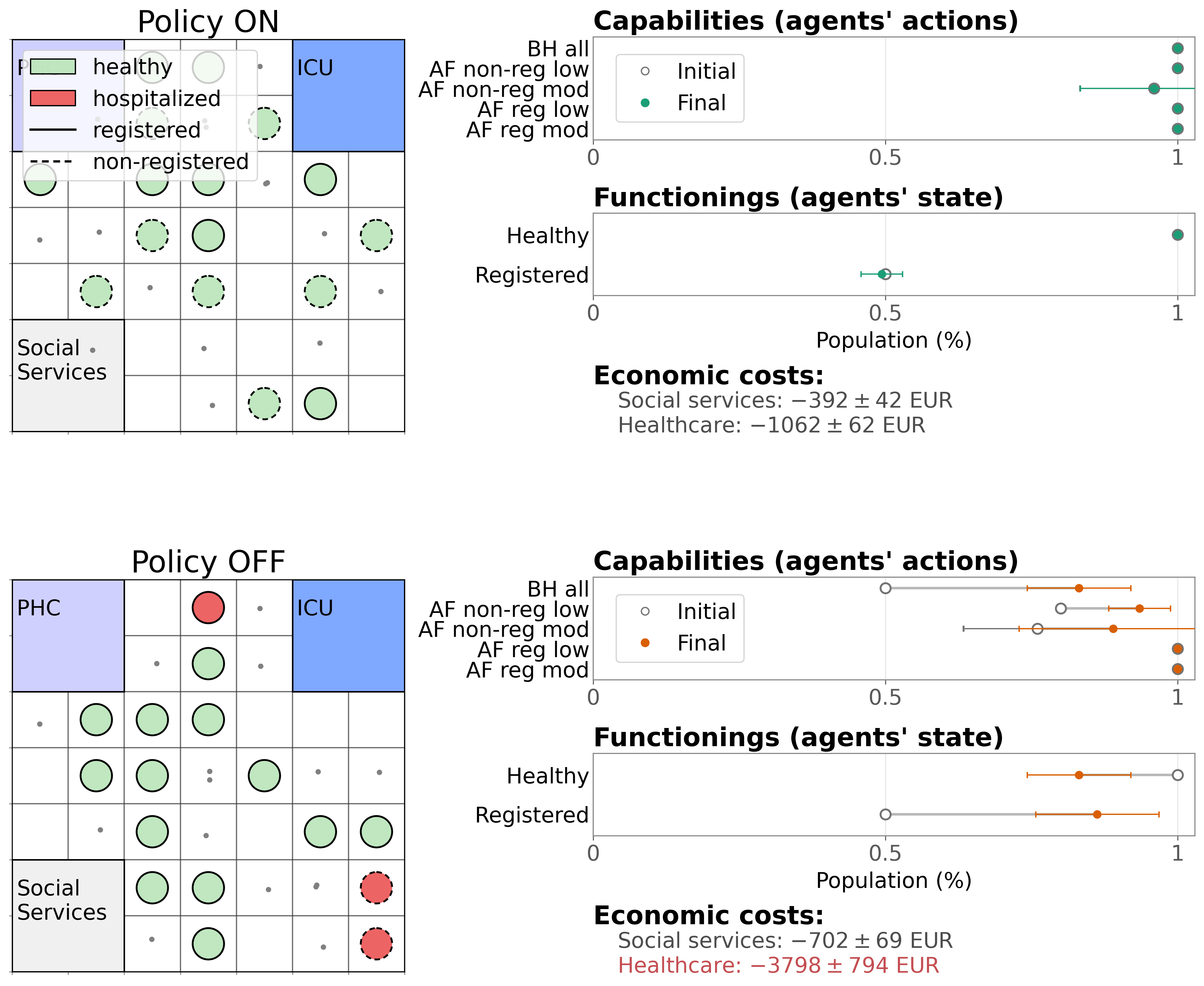}
    \caption{$N =16$ PEH agents}
    \label{fig:8agents}
  \end{subfigure}\hfill
  \begin{subfigure}[b]{0.5\textwidth}
    \includegraphics[width=\textwidth]{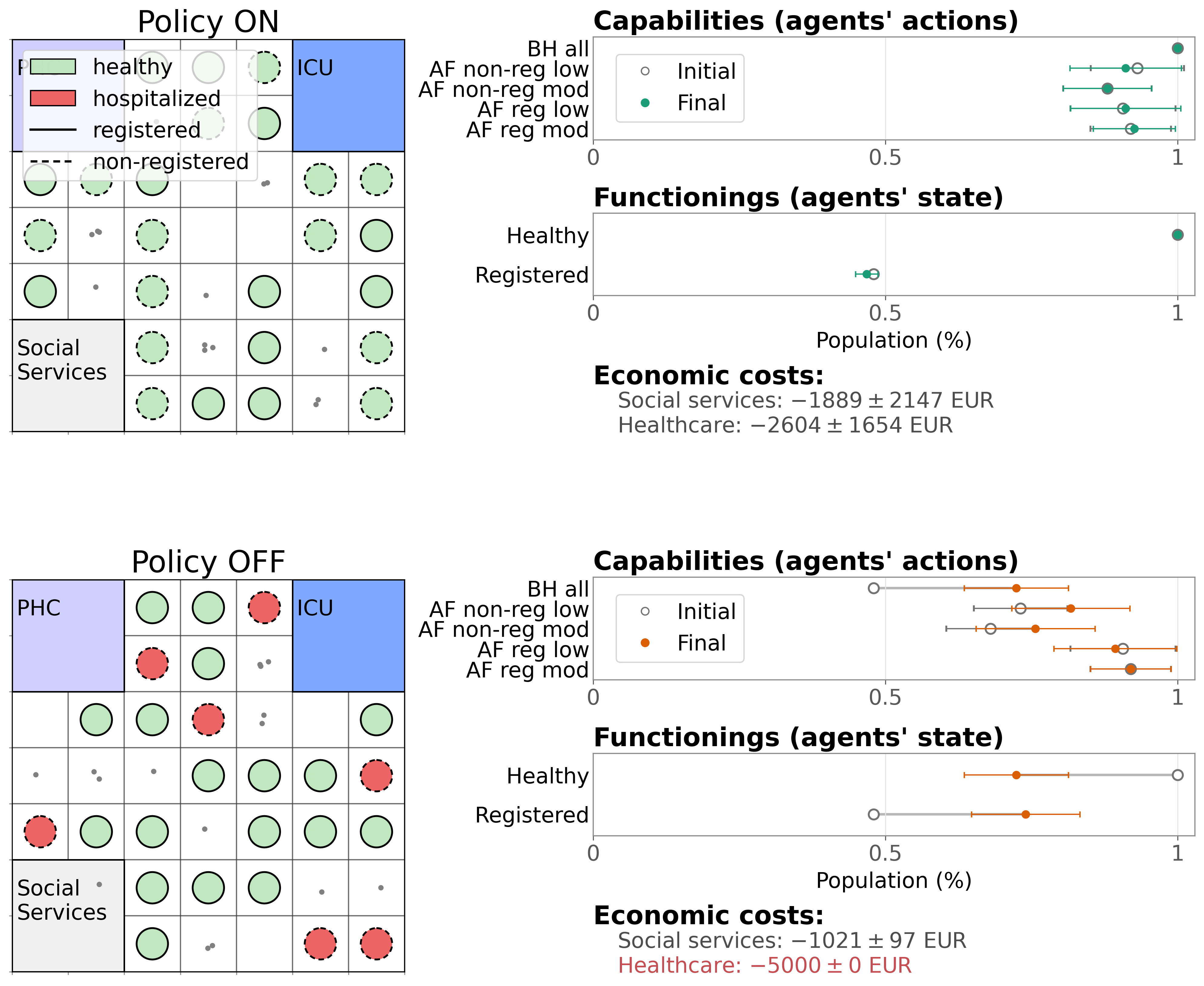}
    \caption{$N=25$ PEH agents}
    \label{fig:16agents}
  \end{subfigure}
    \caption{Comparison of the simulation outcomes with \emph{policy ON} and \emph{policy OFF}
     for $N \in \{16, 25 \}$ PEH agents in (a) and (b). Left: final grid state with social workers' locations (grey smaller dots) and PEH agents' locations (dots coloured by health state and outlined by registration state). Right: Evaluation panel in terms of capabilities (agents' actions), assessed using Eq. (\ref{eq: evaluation}) and abbreviating affiliation and bodily health as \textit{AF} and \textit{BH}, respectively; functionings (agents' state), assessed considering the percentage of agents ending up healthy or hospitalized, and economic costs, assessed considering healthcare and social services expenses, with standard deviations across seeds. }
    \label{fig:finaleval}
\end{figure*}

\section{Evaluating the Impact of Policies on PEH's Healthcare}

The legal policy under evaluation aims to provide inclusive and adapted PHC for PEH, starting by lifting the registration requirement for PHC access (\emph{policy ON}). Without this policy in the system, the current situation requires non-registered PEH agents to engage with social-service workers and complete the ``social registration" (\textit{empadronamiento social}) before accessing PHC (\emph{policy OFF}). Resource limitations are explicit: only a limited number of social workers are available, making trust-building and engagement opportunities limited. Social and healthcare budget are also limited, and we model operational costs based on real ones ($10$ euros for each social worker in the street team, $30$ euros per PHC visit, and $1000$ per hospital admission~\cite{CataloniaHealth2020-OrdreSLT63}). 

We consider an episodic \(7 \times 7\) grid environment with $N_{sw}=15$ social workers, and two different PEH population settings ($N\in \{16, 25\}$). All PEH agents are initialized in a similar level of illness, and their health state decreases linearly by $h=-0.5$ per timestep in the abscence of medical care. Agent profiles are sampled from the synthetic demonstration data $\mathcal{D}$, constructed to match macro-level statistics reported by local non-profits~\cite{ArrelsRecompte2023}. Each PEH agent profile includes attributes such as registration and health status, history life traumatic experience, duration of homelessness, prior encounters with services, age, gender, and income. 

The calibrated parameters $\hat{\theta}$, derived from the calibration procedure (Section~\ref{sec: calibration}), encode profile-dependent behavioral features. In particular, agents with a higher number of prior encounters with social workers, shorter durations of homelessness, and a previous history of traumatic experiences, behave as low-trust agents, which translates into a lower probability of engaging with social services than for moderate-trust agents, who have opposite characteristics. 

We evaluate the effect of the policy on the agents behavior based on both their terminal state (healthy vs.\ hospitalized) and the strategies they follow to reach those outcomes. 
Agents learn these decentralized strategies via independent Q-learning: each agent $i$ learns an $\epsilon$-greedy strategy $\pi_i$ using only its own history of states, actions and rewards, while ignoring the existence of other agents.

 \subsection{Results and Discussion}
Under the previously described conditions, agents find strategies that lead to an optimal healthy status when resources and policy constraints allow them to. 
In Fig~\ref{fig:learning} we analyze the agents' learning when \textit{policy OFF}. Fig.~\ref{fig:rewards} reports the aggregated rewards over learning episodes, where non-registered PEH agents incur into significant negative rewards before finding the optimal strategy after $400$ episodes, while the registered agents converge almost immediately. Importantly, the differences between low- and moderate-trust agents are more visible in Fig.~\ref{fig:policies}, which captures the learned optimal strategy for each subgroup, proven to be empirically robust across random seeds. Moderate-trust agents tend to learn shorter, more direct trajectories toward recovery. In contrast, low-trust agents tend to learn longer strategies, reflecting the need for this complex trust-building process when \emph{policy OFF}. %Their optimal strategies include actions $\hat{a_1}$ and $\hat{a_2}$, as expected for low-trust agents. 

Fig.~\ref{fig:finaleval} provides a snapshot of the last simulation step, comparing \emph{policy ON} vs.\ \emph{policy OFF}. We examine the evolution of central capabilities (bodily health and affiliation), functionings (health and registration states), and system-level indicators (public budgets). All these indicators are systematically lower when \emph{policy OFF}, with much more agents ending in a non-registered and non-healthy state. Additionally, as the PEH population increases with a fixed number of social workers, the agents' engagement opportunities decrease. This reflects a domain-informed point: current larger populations require more social workers on the street. Using our evaluation method, policy-makers could identify clusters of agents who simultaneously lack capabilities and end up with poor functionings, such as the non-registered agents when \emph{policy OFF} (especially non-registered low-trust agents). These agents represent those most vulnerable within the simulation, as they are deprived of both the opportunities and the outcomes necessary for well-being and development. 

Finally, we assess the economic implications under both policy settings. \emph{Policy OFF} produces higher expenditure due to a bigger reliance on emergency services. In contrast, \emph{Policy ON} allows earlier and more direct access to PHC, reducing emergency costs.
With further development, the simulation could help identify the optimal number of street-team social workers, quantify bottlenecks in registration procedures, and evaluate policies holistically across capabilities, functionings, and system-level costs.

\section{Conclusions, Limitations and Future Work}
\label{sec:concl}
In this paper we have presented the first implementation of a simulation framework for real-world social policy-making aligned with the capability approach for human development. The paper responds to the request of non-profit and governmental organizations, to evaluate policies under discussion on health inequity and homelessness. We have developed (i) an ABM environment with explicit policy constraints and limited resources, (ii) RL-based agents calibrated with profile-dependent behavioral heterogeneity, and (iii) a CA-informed evaluation that distinguishes between capabilities (real opportunities) and functionings (realized outcomes). Our findings show how policy constraints impact systemic exclusion from PHC, affecting people experiencing homelessness (PEH) in the city of Barcelona. These results highlight a path to mitigate health inequity not only by providing the necessary material resources, but also by building relationships of trust with PEH that restore their central capabilities and facilitate the engagement of PEH with social services. 

Future work will focus on (i) scaling up to larger populations (e.g., from tens to hundreds of PEH agents) and more realistic spatial representations (e.g., OSM-derived street networks) while checking robustness across seeds; (ii) grounding agent initialization and calibration in richer real-world datasets (not just macrodata), potentially requiring simulation-based calibration; (iii) expanding the PEH agents' state and action sets to model additional policy elements in the system, including the housing dimension (e.g. shelters and other housing initiatives); and (iv) extending the CA-aligned evaluation to a broader set of central capabilities as more actions are introduced. 

\newpage
\section*{Acknowledgments}

This research has been conducted with the support of the EU-funded VALAWAI (\#~101070930) and EVASAI (\#~PID2024-158227NB-C31) projects, as well as the support of the United Nations Economic and Social Commission for Western Asia (UN ESCWA). Special thanks to all the local stakeholders involved. Beatriz Fernández (Fundació Arrels) for sharing the law proposal under discussion and for trusting us to provide insights to the ongoing decision-making. Beatriu Bilbeny (Salutsensellar) for guiding us toward the key topics to address on health inequity, and for contributing to refine the initial model approach. Thanks to Núria Ferran and Bet Bàrbara, for clarifying the functioning of social services and city hall administration in Barcelona. And thanks to the human development studies community, including Flavio Comin, for giving us the necessary feedback on capability representation and measurement.
\bibliographystyle{named}
\bibliography{ijcai26}

\end{document}